\documentclass[journal]{article}

\usepackage{lineno}
\usepackage{latexsym,amssymb,amsmath}
\usepackage{mathabx}
\usepackage{longtable}
\usepackage{tikz,pgf}
\usepackage{subfig}
\usepackage{wrapfig}
\usepackage{rotating}
\usepackage{multirow}
\usepackage{hyperref}
\usepackage{authblk}
\usepackage{natbib}

\usepackage{algorithm}
\usepackage{algorithmic}

\usepackage{graphicx}
\usepackage{epstopdf}
\input{def.set}

\usepackage{url}
\usepackage{hyperref}
\hypersetup{colorlinks,%
citecolor=blue,%
filecolor=blue,%
linkcolor=red,%
urlcolor=blue,%
pdftex}

\begin{document}
%
\title{Integrated multimodal network approach to PET and MRI based on multidimensional persistent homology}

\author[a,d]{Hyekyoung Lee} 
\author[a,e]{Hyejin Kang}
\author[f,g]{Moo K. Chung}
\author[c]{Seonhee Lim}
\author[b]{Bung-Nyun Kim}
\author[a,d]{Dong Soo Lee} 



\affil[a]{Department of Nuclear Medicine,}
\affil[b]{Division of Child and Adolescent Psychiatry, Seoul National University College of Medicine,}
\affil[c]{Department of Mathematical Sciences,}
\affil[d]{Department of Molecular Medicine and Biopharmaceutical Sciences, Graduate School of Convergence Science and Technology, and College of Medicine or College of Pharmacy, }
\affil[e]{Data Science and Knowledge Creation Research Center, Seoul National
University, \\ Seoul, Korea.}
\affil[f]{Department of Biostatistics and Medical Informatics,}
\affil[g]{Waisman Laboratory for Brain Imaging and Behavior, University of Wisconsin, Madison,
WI, U.S.A.}

\maketitle

\begin{abstract} 
Finding the 
underlying 
relationships among multiple imaging modalities in a coherent fashion is one of challenging problems in the multimodal analysis.
In this study, we propose a novel multimodal network approach based on multidimensional persistent homology.
In this extension of the previous threshold-free method of persistent homology,
we visualize and discriminate the topological change of integrated brain networks by varying not only threshold but also mixing ratios between two different imaging modalities.
Moreover, we also propose an integration method 
for multimodal networks, called 
one-dimensional projection, with 
a specific mixing ratio between modalities.
We applied the proposed methods to PET and MRI data 
from 23 attention deficit hyperactivity disorder (ADHD) children, 21 autism spectrum disorder (ASD), and 10 pediatric control subjects.
From the results, we found that the brain networks of ASD, ADHD children and controls 
differ significantly, with ASD and ADHD showing asymmetrical changes of connected structures between metabolic and morphological connectivities.  
During the integration of PET and MRI, ASD children showed stronger connections than controls in the metabolic connectivity, but weaker connections in the morphological connectivity than controls. 
On the other hand, ADHD had different connected structures only in the morphological connectivity compared to controls. 
These results provide a multidimensional homological understanding of disease-related 
PET and MRI networks 
that discloses the network association with 
ASD and ADHD.
\end{abstract}


%

\section{Introduction} 

Noninvasive brain imaging techniques such as fluorodeoxyglucose (FDG) positron emission tomography (PET) and T1-weighted magnetic resonance imaging (MRI) disclose different characteristics of the human brain.
PET reveals regional brain metabolism and MRI the brain morphology \citep{bassett.2008.jneuro,gong.2009.jneuro,he.2007.cc,phelps.1998.aim}.
The inter-subject and inter-regional correlation of brain metabolic uptake between brain regions on FDG PET or brain morphology on MRI could model the brain metabolic or morphologic networks, respectively \citep{bernhardt.2011.cc,chen.2008.cc,chung.2013.miccai,he.2007.cc,hosseini.2012.plosone,huang.2010.ni,lee.2008.ejnmmi,toussaint.2012.ni}.
We 
refer to these correlation networks based on FDG PET as metabolic networks and those based on T1 MRI as morphological networks, both of which 
differ from the functional network on functional MRI or structural network on diffusion tensor imaging (DTI).
The challenge, however, remains to find 
a way 
to integrate the networks acquired from the two different imaging modalities of PET and MRI.

The simplest way to integrate two different weighted networks 
is to find common or coincidental significant connections by %
performing a parallel statistical analysis of each modality \citep{honey.2009.pnas,vanheuvel.2009.hbm}.
This approach 
works when 
one tries to compare the results from each modality within a group or between groups; however, it 
is difficult to find the common-but-hidden or discordant connections.
The other approach 
is to construct the integrated network using multimodal imaging data by overlapping the connections from each modality or by weighting anatomical connectivity to a functional one \citep{bowman.2012.ni,hosseini.2013.ni}.
These methods 
are disadvantageous because it is necessary to determine the proper threshold a priori for each network as well as to choose the mixing ratio of the two in an appropriate way.
Bowman, et. al. (2012) tried to minimize the proposed objective function related with clustering performance,
whereas Hosseini, et. al. (2013) showed the changes of topological measures of the integrated networks at all network densities \citep{bowman.2012.ni,hosseini.2013.ni}. 

In this paper, we propose a new analytical framework based on multidimensional persistent homology that combines networks of two different imaging modalities.
We first observed all the changes of topological structure of multimodal integrated networks without fixing thresholds and with varying mixing ratios of the two modalities and then looked for the integrated network having 
a mixing ratio with significant discrimination between the disease group and the controls after
performing a one-dimensional projection.
The proposed method is a multimodal approach that 
extends the concept of network filtration to two (multiple) dimensions \citep{lee.2012.ieeemi}.

We previously proposed
the concept of network filtration based on persistent homology 
to solve the thresholding problem of unimodal network analysis \citep{lee.2012.ieeemi}.
Whereas we vary single type of threshold in unidimensional persistent homological analysis, 
multidimensional persistent homology allows two or more thresholds for multimodalities \citep{carlsson.2009.dcg}.
The multifiltration method based on bi-dimensional persistent homology
allowed 
integration of two different weighted networks into the bi-sequence of unweighted networks 
as the thresholds 
are varied simultaneously.
We estimated topological invariants from each modality and visualized their changes during multifiltration.
Here we used the number of connected components (CCs), called the zeroth Betti number $\beta_{0},$ as the topological invariants.

In the bi-sequence of unweighted networks, we could extract the sequence of unweighted networks along a projection line with specific mixing ratio between metabolic and morphological networks.
This procedure is
a one-dimensional projection of multifiltration.
Along the projection line of the sequences of unweighted networks, we could reorder the edges, reestimate their weights, and construct a new integrated 
multimodal network.
The change 
in the connected structure of
the integrated network could then be represented 
in an algebraic form, known as
a single linkage matrix (SLM).
A group comparison of
the $\beta_{0}$-plot was 
performed by
using a Kolmogorov-Smirnov-like (KS-like) test and
deriving a one-dimensionally projected SLM by
using Gromov-Hausdorff (GH) distance, both of which were based on 
permutation methods \citep{chung.2013.miccai,lee.2012.ieeemi}.

Our main contributions over the previous publications including our own work \citep{lee.2012.ieeemi} are
as follows:
\begin{enumerate}
\item We devise a multivariate extension of univariate persistent homology and its application to multimodal brain network analyses of PET and MRI;
\item we develop 
a new visualization tool,
the $\beta_{0}$-plot, for showing the changes of the integrated connected structures of metabolic to morphological correlation networks with various mixing ratios; and
\item we extract 
the integrated network of metabolic and morphological 
networks at a certain mixing ratio, discriminating the disease group from the controls by
using a one-dimensional projection and its representation as an SLM.
\end{enumerate}
In our study, the dataset consisted of FDG PET and T1 MRI
images of 23 attention deficit hyperactivity disorder (ADHD), 21 autism spectrum disorder (ASD) children and 10 pediatric control subjects.
We used 93 regions of interest (ROIs) based on
an automated anatomical labeling atlas (AAL) as the nodes of networks \citep{tzourio-mazoyer.2002.ni}.
The mean FDG uptake and average Jacobian determinant within
the ROIs are the measured indices in this article for PET and MRI, respectively, which 
were later used to generate
the networks.
After constructing the correlation matrix based on inter-subject variability using PET and MRI, we applied the proposed methods, multifiltration, and its one-dimensional projection.

\section{Materials and methods}
\label{sec:methods}

\subsection{Network construction}
\label{sec:network_construction}

The $93$ ROIs serve as nodes, $V = \left\{ v_{1}, \dots, v_{p} \right\}$ $(p=93).$
PET and MRI data have the identical node set $V$ in the same template space.
On each node $v_{i},$ we have two different imaging measurements $\bu_{i}^{P} \in \Real^{n \times 1}$ and $\bu_{i}^{M} \in \Real^{n \times 1}$ obtained from PET and MRI, respectively.
If $\bu_{i} \in \left\{ \bu_{i}^{P}, \bu_{i}^{M} \right\}$ is assumed to be normally distributed with the mean $0$ and the variance $1/n,$ the distance between $v_i$ and $v_j$ is determined by one minus 
Pearson correlation, 
$1 - \mbox{corr}(\bu_{i},\bu_{j}).$
$x_{ij}$ and $y_{ij}$ are written as a distance between $v_{i}$ and $v_{j}$ of PET and MRI, respectively. 
We denote the weighted networks for PET and MRI as $\calP(V,\bX)$ and $\calM(V,\bY).$  where  $\bX=[x_{ij}]$ and $\bY=[y_{ij}]$ are the distance matrices of the PET and MRI networks, respectively.


\subsection{Multidimensional persistence}
\label{sec:multifiltration}

The persistent homology has been introduced to solve the thresholding problem of unimodal brain network analysis \citep{lee.2012.ieeemi}.
Given one weighted network $\calP(V,\bX)$ and threshold $\epsilon,$  a unweighted network $B_{\calP} (\epsilon)$ is obtained by filtering the weighted network $\calP(V,\bX)$ by the  threshold or filtration value $\epsilon.$
If the weighted network is repeatedly filtered for the ordered thresholds 
$\left\{ \epsilon_{\min} = \epsilon_{1} < \epsilon_{2} < \dots < \epsilon_{q} = \epsilon_{\max} \right\},$
it is decomposed into the sequence of unweighted networks which satisfy the nested property: 
$$B_{\calP} (\epsilon_{1}) \subseteq B_{\calP} (\epsilon_{2}) \subseteq \cdots \subseteq B_{\calP} (\epsilon_{q}).$$
This procedure is called a graph filtration \citep{giusti.2015.pnas,lee.2011.miccai,lee.2012.ieeemi}. 

Here we extend this filtration method to the multidimensional version by introducing a multidimensional persistence  \citep{carlsson.2009.dcg}.
Suppose that two weighted networks $\calP(V,\bX)$ and $\calM(V,\bY)$ are given. 
They share a common node set, but have different distance matrices between nodes. 
Two weighted networks are simultaneously bi-filtered at two thresholds $\omega$ and $\upsilon$ {\em via} 
\be
B_{\calP,\calM} (\omega, \upsilon) = B_{\calP} (\omega) \cap B_{\calM} (\upsilon). 
\label{eq:and}
\ee
The bi-filtered unweighted network $B_{\calP,\calM} (\omega, \upsilon)$ is obtained by connecting edges that satisfy $x \le \omega$ and $y \le \upsilon$ in $B_{\calP} (\omega)$ and $B_{\calM} (\upsilon),$ respectively. 
If the filtration values are given 
by $\omega_{1} < \omega_{2} < \dots < \omega_{q}$ and $\upsilon_{1} < \upsilon_{2} < \dots < \upsilon_{q},$ the multifiltration can be written as
\bee
\small{
\begin{array}{ccccccc}
B_{\calP,\calM} (\omega_{1},\upsilon_{1}) & \rightarrow & \dots & \rightarrow & B_{\calP,\calM} (\omega_{q},\upsilon_{1}) \\
\downarrow& & & & \downarrow \\
\vdots & & \ddots & & \vdots \\
\downarrow& & & & \downarrow \\
B_{\calP,\calM} (\omega_{1},\upsilon_{q}) & \rightarrow & \dots & \rightarrow & B_{\calP,\calM} (\omega_{q},\upsilon_{q}).
\end{array}}
\eee
The bi-filtration also satisfies the nested property:
\be
\label{eq:multi_persistence}
B_{\calP,\calM} (\omega_{i},\upsilon_{l}) \subseteq B_{\calP,\calM} (\omega_{j},\upsilon_{m}) \mbox{ for } \omega_{i} \leq \omega_{j} \mbox{ and } \upsilon_{l} \leq \upsilon_{m}.
\ee

In the Algebraic Topology, the Betti number is used to determine the shape of topological spaces including networks and to distinguish topological spaces \citep{adler.2010.arxiv,carlsson.2005.ijsm,edelsbrunner.2008.cm,ghrist.2008.bams}. 
The zeroth Betti number $\beta_{0}$ is the number of CCs which are subsets of the network, where any nodes are connected through edges. 
In this study, we choose $\beta_{0}$ as the topological invariant and estimate them from the obtained bi-sequence of unweighted networks during the bi-filtration. 
The change of $\beta_{0}$ during filtration is usually visualized by the barcode \citep{carlsson.2009.dcg}. 
However, since the barcode visualizes the change of CCs using bars when varying a threshold, it is not proper to represent the change of $\beta_{0}$ with respect to two different thresholds $\omega$ and $\upsilon.$
Thus, we use $\beta_{0}-$plot which visualizes the change of number of CCs with respect to two thresholds $\omega$ and $\upsilon.$  

\begin{figure*}[t]
\begin{center}
\includegraphics[width=1\linewidth]{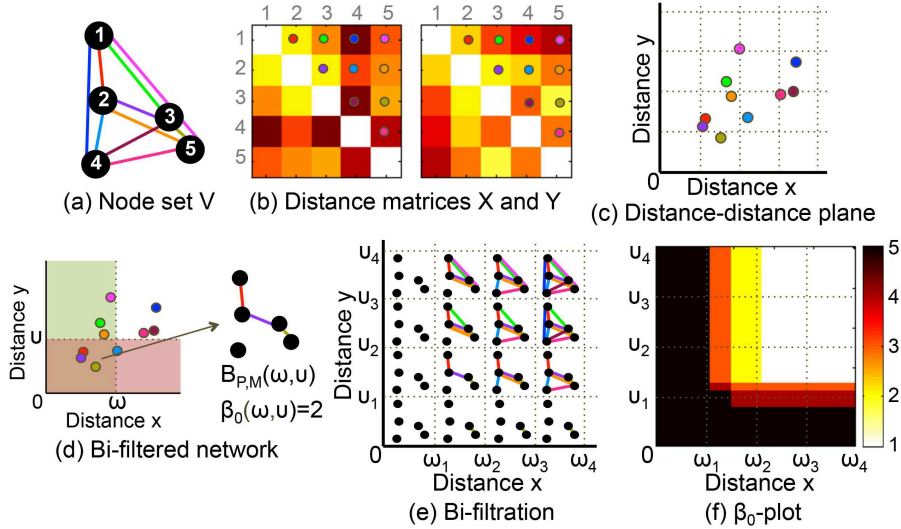} 
\caption{(a,b) Two networks $\calP(V,\bX)$ and $\calM(V,\bY)$ are given with a node set $V$ and  distance matrices $\bX=[x_{ij}]$ and $\bY=[y_{ij}].$ (c) The distance pair $(x_{ij},y_{ij})$ of the edge connecting nodes $i$ and $j$ is plotted on the x (horizontal) axis and y (vertical) axis of the x-y plane. The color of dot indicates which edge comes from (a) and (b).  
(d) The bi-filtered unweighted network $B_{\calP,\calM} (\omega,\upsilon)$ is constructed by connecting edges that satisfy both $x \le \omega$ and $y \le \upsilon.$
In this bi-filtered network, the number of CCs $\beta_{0}$ at $(\omega,\upsilon)$ is $2.$
(e) Bi-sequence of unweighted networks is obtained by bi-filtration of $\calP(V,\bX)$ and $\calM(V,\bY).$ (f) The number of CCs $\beta_{0}$ of (e) is plotted on the distance(x)-distance(y) plane. This is called $\beta_{0}$-plot.}
\label{fig:example_multifiltration}
\end{center}
\end{figure*}

The example of bi-filtration is shown in Fig. \ref{fig:example_multifiltration}. 
Two networks $\calP(V,\bX)$ and $\calM(V,\bY)$ share nodes, but have different distance measures in (a) and (b). 
Each edge is encoded into a distance pair $(x_{ij},y_{ij})$ and plotted on the distance-distance(DD) domain in (c). 
When two networks are simultaneously bi-filtered at the threshold pair $(\omega,\upsilon),$ only edges in the region $x \le \omega$ $\&$ $y \le \upsilon$ are connected as shown in (d). 
The bi-sequence of unweighted network is obtained by bi-filtration at $(\omega_{1},\upsilon_{1}), \dots, (\omega_{4},\upsilon_{4})$ in (e). 
Its $\beta_{0}-$plot is illustrated in (f). 
The $\beta_{0}$ is a decreasing function of $(\omega,\upsilon)$ with the range from $1$ to $p.$

\subsection{One-dimensional projection}
\label{sec:onedim_proj}

\begin{figure*}[t]
\begin{center}
\includegraphics[width=1\linewidth]{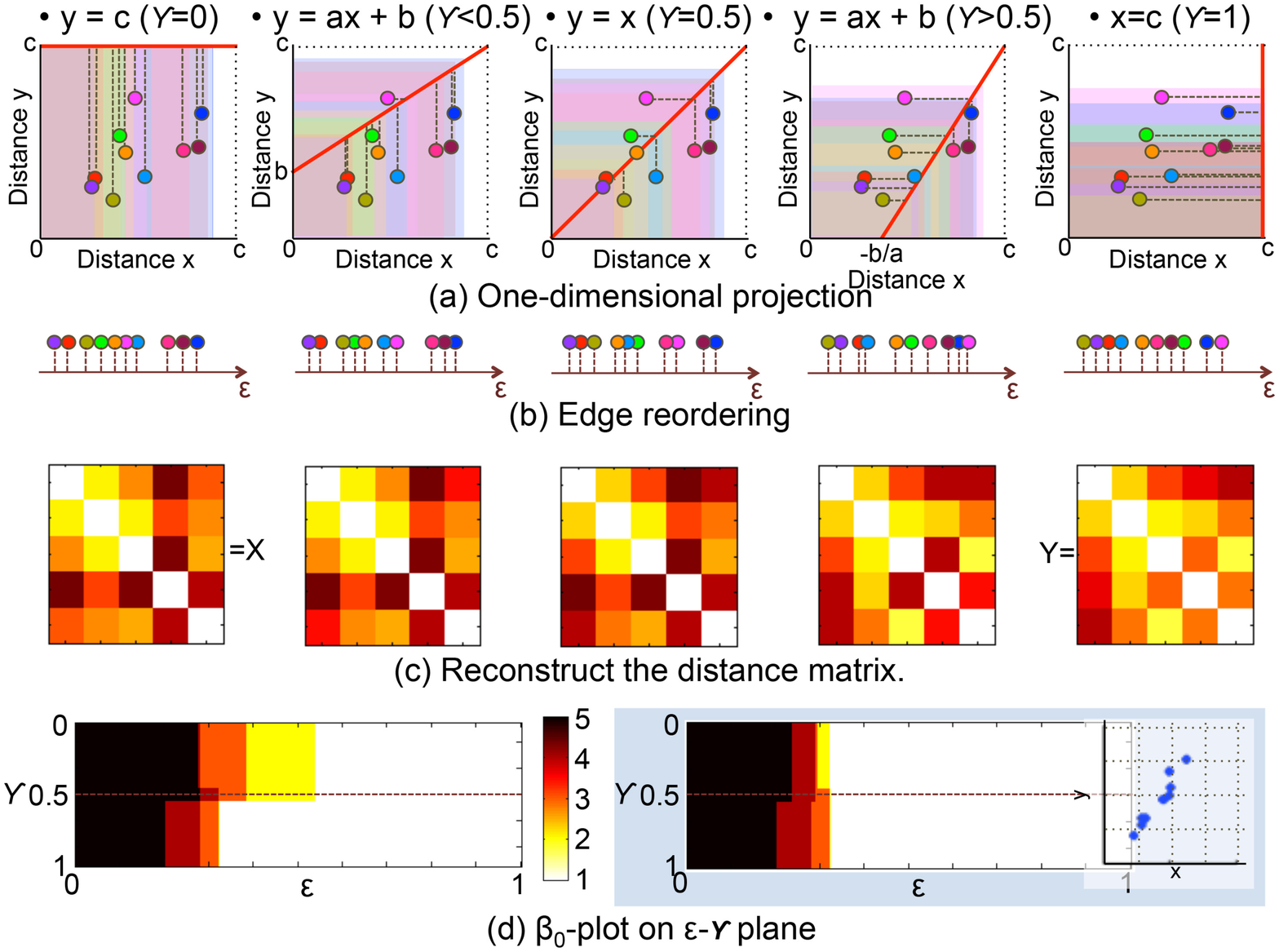} 
\caption{One-dimensional projection of the example in Fig. \ref{fig:example_multifiltration}. (a) The edges are projected onto the line $y=ax+b$ in (\ref{eq:line}) when $\gamma=0,0.4,0.5,0.6,$and $1$ from left to right. (b) Reordered edges along the projected line. The edge weight (distance) $\epsilon$ is recalculated on the projected line. (c) According to the new edge weights, an edge weight matrix is reconstructed at each $\gamma.$ The reconstructed matrices at $\gamma=0$ and $1$ are exactly the same to $\bX$ and $\bY$ in Fig. \ref{fig:example_multifiltration} (b).
(d) The $\beta_{0}-$plot on the $x-y$ plane in Fig. \ref{fig:example_multifiltration} (f) is transformed onto the $\epsilon-\gamma$ plane. As shown in the right blue box, if two kinds of edge weights are similar, $\beta_{0}-$plot on the $\epsilon-\gamma$ plane tends to be symmetric with respect to $\gamma=0.5.$}
\label{fig:example_one_dimensional_projection}
\end{center}
\end{figure*}

The filtration is a procedure to add edges in increasing order of edge weights.
When two different kinds of distance (or edge weight) measures are defined in the network, the order of adding edges depends upon how to prioritize two measures.  
Here we change the priority of two measures by controlling a mixing ratio $\gamma$ between them and increase a pair of thresholds $(\omega,\upsilon)$ along the line with mixing ratio    
\be
y = ax + b,  \quad \left( a = {\scriptstyle{\frac{\gamma}{1-\gamma}}},  \mbox{   } b = {\scriptstyle{\frac{1-2\gamma}{1-\gamma}}}c \right)
\label{eq:line}
\ee
where $\gamma$ is in $[0,1]$ and $c$ is a large enough positive constant satisfying $x,y \in [0,c].$
If we choose $\gamma=0,$ two thresholds $(\omega,\upsilon)$ are varied along the line $y=c$ and the order of edges is affected only by $\omega$ because $\upsilon=c$ and $\left\{(x,y) | x \le \omega \mbox{ } \& \mbox{ } y \le c \right\}=\left\{(x,y) | x \le \omega \right\}.$
If we choose $\gamma=1,$ the order of edges is affected only by $\upsilon.$
The reason why the equation of the line (\ref{eq:line}) is somewhat complicated is to include such uni-filtered cases.
If $0< \gamma < 0.5,$ the procedure to add edges is more influenced by the change of $\omega$ under the condition of $\upsilon \ge b.$ 
If $0.5 < \gamma < 1,$ it is more influenced by the change of $\upsilon$ under the condition of $\omega \ge -b/a.$ 
If $\gamma = 0.5,$ it is affected equally by $\omega$ and $\upsilon.$ 

Suppose that two sequence of thresholds $0 \leq \omega_{1} \leq \dots \leq \omega_{q} \leq c$ and $0 \leq \upsilon_{1} \leq \dots \leq \upsilon_{q} \leq c$ are given, and $\omega_{i}$ and $\upsilon_{i}$ satisfy the equation (\ref{eq:line}) with the given $\gamma.$
The filtration along the line generates a sequence of unweighted networks as follows:
\bee
B_{\calP,\calM} (\omega_{0},\upsilon_{0}) \rightarrow B_{\calP,\calM} (\omega_{1},\upsilon_{1}) \rightarrow \cdots \rightarrow  B_{\calP,\calM} (\omega_{q},\upsilon_{q}).
\eee
This procedure is called one-dimensional projection and the projected sequence of bi-filtered networks satisfies the persistence property in Eq. (\ref{eq:multi_persistence}). 
We integrate two different kinds of measures of edge weight by one-dimensional projection and control the integration ratio by $\gamma.$ 
The example of one-dimensional projection is shown in Fig. \ref{fig:example_one_dimensional_projection} when $\gamma=0, < 0.5, =0.5, > 0.5,$ and $ =1.$ 

One-dimensional projection can be though as a function $\pi$ that project the edge onto the line (\ref{eq:line}) as follows:   
\bee
\pi : (x_{ij},y_{ij}) \rightarrow (x_{ij}',y_{ij}')= \left\{ \begin{array}{ll} 
\left(\frac{y_{ij}-b}{a},y_{ij} \right)& \mbox{if } y_{ij} > ax_{ij} + b, \\
\left(x_{ij},ax_{ij} +b \right)& \mbox{otherwise. }
\end{array} \right.
\eee
A new edge weight $z_{ij}$ on the projected line is the normalized Euclidean distance between x-intercept when $\gamma > 0.5$ or y-intercept when $\gamma \le 0.5$ and $(x_{ij}',y_{ij}'), $ that is, 
\be
z_{\gamma,ij} = \left\{ \begin{array}{ll}
\frac{\sqrt{{x'}_{ij}^{2} + ({y}_{ij}'-b)^{2}}}{\sqrt{c^{2} + (c-b)^{2}}} & \mbox{if } \gamma < 0.5, \\ 
\frac{\sqrt{({x'}_{ij}+\frac{b}{a})^{2} + {y'}_{ij}^{2}}}{\sqrt{(c+\frac{b}{a})^{2} + c^{2}}} & \mbox{otherwise}.
 \end{array} \right.
\label{eq:newweight}
\ee
The denominator of (\ref{eq:newweight}) is a normalization term to set the maximum filtration value as $1.$
In this way, an integrated network with a new edge weight matrix $\bZ_{\gamma}=[z_{\gamma,ij}]$ is obtained at each $\gamma$ by reweighing and reordering edges on the projected line. 
We can apply the filtration to the new integrated network and estimate the change of $\beta_{0}$ by varying a new threshold $\epsilon.$
The proposed method 
of multifiltration
contains all possible edge-sorting procedures with various mixing ratios $\gamma$ between two kinds of edge weights.

Fig. \ref{fig:example_one_dimensional_projection} shows a procedure of one-dimensional projection of the example in Fig. \ref{fig:example_multifiltration}. 
In (a), the edges are projected onto the lines with $\gamma = 0, 0.4,0.5,0.6,$ and $1$ from left to right.  
The reordered edges on the projected line are shown in (b) and new integrated edge weight matrices estimated at various $\gamma$s are shown in (c).  
After the filtration of each edge weight matrix, the number of CCs $\beta_{0}(\epsilon,\gamma)$ is plotted on the $\epsilon-\gamma$ plane in the left part of (d). 
The figure in (d) has the same information of $\beta_{0}$ as Fig. \ref{fig:example_multifiltration} (f), but the former is plotted on the $\epsilon-\gamma$ plane and the latter is on the $x-y$ plane.  
The right part in the blue box of (d) shows another example of DD plot of $10$ edges. 
In this example, two different edge weight measures are highly correlated. 
In this case, the $\beta_{0}-$plot is more symmetric with respect to $\gamma=0.5$ than one in the left part of (d).

\subsection{$\beta_{0}$-plot comparison}
\label{sec:barcode_comparison}

Chung, et el. (2013) proposed Kolmogorov-Smirnov-like  (KS-like) test statistic, $T = \sup_{\epsilon} | \beta_{0}^{1} (\epsilon) - \beta_{0}^{2} (\epsilon) | $ for testing the difference between two barcodes $\beta_{0}^{1}$ and $\beta_{0}^{2}$ obtained from two different groups \citep{chung.2013.miccai}.
KS-like test statistic can also be applied to multidimensional $\beta_{0}$-plot comparison.
In our case, $\beta_{0}$ is a function of two filtration values $(\epsilon,\gamma)$ instead of one filtration value $\epsilon:$
$$T = \sup_{\epsilon,\gamma} | \beta_{0}^{1} (\epsilon,\gamma) - \beta_{0}^{2} (\epsilon,\gamma) |. $$
The KS-like test statistic for testing the null hypothesis that two $\beta_{0}$-plots were not different is the maximum of absolute value of $\beta_{0}^{1} (\omega,\upsilon) - \beta_{0}^{2} (\omega,\upsilon).$
Since we are maximizing over all possible thresholds $\epsilon$ and $\gamma$, the multiple comparison issues are automatically taken care of.
The null distribution is estimated by the permutation method.

We also check the symmetry of $\beta_{0}-$plot on the $\epsilon-\gamma$ plane with respect to $\gamma=0.5.$
The symmetry $\delta$ is estimated by 
$$\delta = \frac{1}{0.5} \int_{0}^{1} \int_{0}^{0.5} |\beta_{0}(\epsilon,\gamma) - \beta_{0}(\epsilon,1-\gamma)| d\gamma d\epsilon.$$ 
The more symmetrical the $\beta_{0}-$plot is, the closer $\delta$ is to $0.$
If two edge weight matrix are exactly the same, their $\beta_{0}-$plot is symmetric and its $\delta$ is equal to $0.$ 
Although the reverse is not true, we can compare global connected patterns between two edge weight matrices using the symmetry of their $\beta_{0}-$plot.
Here, we used the symmetry of $\beta_{0}-$plot in two ways. 
First, we integrated two different modalities, PET and MRI, using $\beta_{0}-$plot and compared global connected patterns between PET and MRI in each group. 
Second, we applied the $\beta_{0}-$plot method to the integrated edge weight matrices of each pair of groups and compared global connected patterns between ADHD and ASD, ADHD and CON, and ASD and CON at various mixing ratio $\gamma.$

\subsection{Single linkage matrix and Gromov-Hausdorff distance} 

The single linkage matrix (SLM) $\bD$ shows the local change of CCs \citep{carlsson.2008.preprint,carlsson.2010.jmlr,lee.2012.ieeemi}.
The element of SLM $\bD=[d_{ij}]$ is the minimum filtration value when two nodes $v_{i}$ and $v_{j}$ are connected directly or indirectly by merging into the same CC.
The mathematical definition is given by
\bee
\label{eq:sld}
\bD = [d_{ij}] = [\min_{P_{ij}} \max_{l} x_{p_{l},p_{l+1}}],
\eee
where $P_{ij} = \{ v_i = p_0, \cdots, p_{k} = v_j \}$ is a path between two nodes $v_i$ and $v_j$ \citep{gower.1969.as}.
The minimum is taken over every possible path $P_{ij}$ between $v_i$ and $v_j$.
The quantity $d_{ij}$ is called a single linkage distance (SLD).

The Gromov-Hausdoff (GH) distance was used for estimating the difference between two matrices such as edge weight and SLMs \citep{carlsson.2008.preprint,carlsson.2010.jmlr,lee.2012.ieeemi}.
The GH distance 
can be viewed as a special case of general framework of 
type-I-error
estimation under multiple comparisons \citep{chung.2013.book}.
Given two SLMs $\bD^{1}=[d^{1}_{ij}]$ and $\bD^{2}=[d^{2}_{ij}]$ with the same node set $V,$ their difference is found by the hypothesis test by setting up a null hypothesis of no difference between two edge weights $d^{1}_{ij}$ and $d^{2}_{ij}$, and an alternative hypothesis:
\be
H^{B}_{0} : d^{1}_{ij} = d^{2}_{ij} \mbox{ v.s. }
H^{B}_{1} : d^{1}_{ij} \neq d^{2}_{ij}.
\label{eq:null_pairwise}
\ee
This hypothesis is for local difference of the connectivity $d^{1}_{ij}$ and $d^{2}_{ij}$ in two different networks.
For the comparison of global difference between two different networks, the hypotheses are given
by
\be
H_{0} : d^{1}_{ij} = d^{2}_{ij} \mbox{ for all } i,j \mbox{ v.s. }
H_{1} : d^{1}_{ij} \neq d^{2}_{ij} \mbox{ for some } i,j.
\label{eq:null_all}
\ee
The null hypothesis $H_{0}$ is the intersection of collection of hypotheses \citep{chung.2013.book}
$$
H_{0} = \bigcap_{\forall i,j} H^{B}_{0}(i,j).
$$
Then, the type-I error $\alpha$ for testing two sided test under the multiple comparisons is given
by
\be
\alpha &=& P \left( \bigcup_{\forall i,j} \left\{ Z(i,j) > h \right\} \right) \nonumber
= 1 - P \left( \bigcap_{\forall i,j} \left\{ Z(i,j) \le h \right\} \right) \\
&=& 1- P \left( \sup_{\forall i,j} Z(i,j) \le h \right)
= P \left( \sup_{\forall i,j} Z(i,j) > h \right),
\label{eq:alpha}
\ee
where $Z(i,j)$ is actually $|d^{1}_{ij} - d^{2}_{ij}|$, i.e., the difference between SLDs.

The Gromov-Hausdorff (GH) distance between two edge weight matrices is defined as
$$d_{GH}(\bD^{1},\bD^{2}) = \sup_{\forall i,j} |d^{1}_{ij} - d^{2}_{ij}|$$
\citep{carlsson.2008.preprint,lee.2012.ieeemi,lee.2011.miccai,memoli.2011.caip}.
$d_{GH}$ can substitute $\sup_{\forall i,j} Z(i,j)$ in (\ref{eq:alpha}).
Now, we can write the type-I error of group differences through GH distance as
$$\alpha = P \left( d_{GH} (\bD^{1},\bD^{2}) > h \right).$$

The permutation method is performed to 
test (\ref{eq:null_all}).
Two bi-modal datasets of PET and MRI for group $1$ and group $2$ with sample size $n_{1}$ and $n_{2}$ are given. 
The group labels are shuffled randomly and two distance matrices $\bX^{i}$ and $\bY^{i}$ of PET and MRI networks are estimated for the $i$th group $(i=1,2).$
By applying one-dimensional projection with $\gamma=0,\dots,1,$ we obtain the sequence of edge weight matrices $\frac{1}{2}\bX^{i}=\bZ^{i}_{0}, \dots,\bZ^{i}_{1}=\frac{1}{2}\bY^{i}$ by the equation (\ref{eq:newweight}) and one of SLMs $\bD^{i}_{0}, \dots,\bD^{i}_{1}.$
This procedure is repeatedly done $5000$ times.
Then, we estimate the null distribution of $d_{GH}(\bD^{1}_{\gamma}, \bD^{2}_{\gamma})$ at $\gamma=0,\dots,1.$
Using two original datasets, we estimate $d_{GH}(\bD^{1}_{true,\gamma},\bD^{2}_{true,\gamma})$.
Then, the type-I error $\alpha$ with the hypothesis (\ref{eq:null_all}) for global difference
is calculated by the percentile of $d_{GH}(\bD^{1}_{true,\gamma},\bD^{2}_{true,\gamma})$ in the null distribution of $d_{GH}(\bD^{1}_{\gamma},\bD^{2}_{\gamma}).$

\section{Results}

\subsection{Subjects and image preprocessing}
\label{sec:preprocessing}

We used FDG PET and MRI data sets: 23 ADHD children (mean age = $8.1 \pm 1.6$ years), 21 ASD children (mean age = $6.0 \pm 1.6$ years) and 10 control subjects (mean age = $9.5 \pm 2.6$ years).
The ADHD children were diagnosed by DSM-IV diagnostic criteria, Korean version of ADHD rating scale IV (K-ARS) and, Korean version of Kiddie-Schedule for Affective Disorders and Schizophrenia-Present and Lifetime version (K-SADS-PL). The ASD children were diagnosed by the Korean version of the Autism Diagnostic Interview-Revised (K-ADI-R) and the Korean version of the Autism Diagnostic Observation Schedule (ADOS).
The control data was obtained from 10 children who failed to meet the criteria of psychiatric disorder or visited for IQ evaluation.
This study was approved by the Institutional Review Board of Seoul National University College of Medicine.

PET images were obtained by ECAT EXACT 47 PET scanner (Siemens-CTI, Knoxville, TN).
They were preprocessed using the statistical parametric mapping toolbox (SPM) \citep{friston.1995.hbm}.
The brain was parcellated into 93 regions of interest (ROIs) based on AAL \citep{tzourio-mazoyer.2002.ni}.
We redivided $16$ cerebellar regions in AAL into three regions, left, right and vermis.
The mean FDG uptake within $93$ ROIs was extracted as a measurement of PET.
The MRI data was segmented by the customized pediatric templates in Template-O-Matic toolbox \citep{wilke.2008.ni}.
All gray matter MRIs were transformed and smoothed and the Jacobian determinant maps were computed based on VBM8 toolbox for SPM.
The mean Jacobian value within $93$ ROIs was extracted as a measurement of MRI using MarsBar toolbox (\href{http://marsbar.sourceforge.net}{http://marsbar.sourceforge.net}).
We factored out the effects of age in PET and MRI using the general linear model (GLM).

\subsection{Distance-distance plot} 

\begin{figure}[tph]
\centering
\includegraphics[width=1\linewidth]{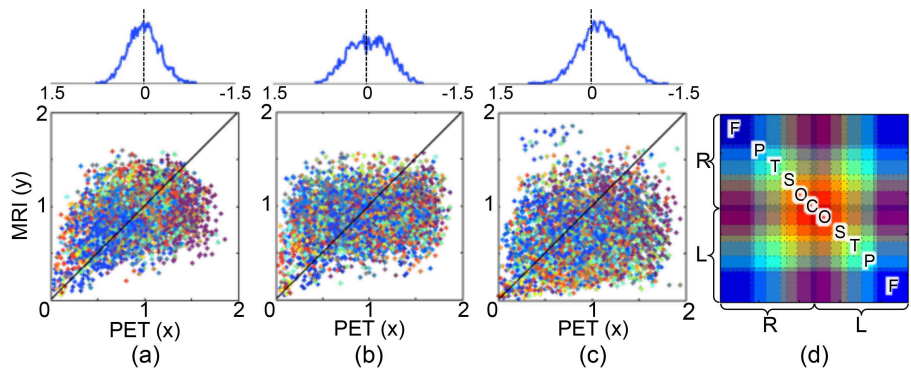} 
\caption{The distance-distance (DD) plot of (a) ADHD, (b) ASD and (c) CON. The horizontal and vertical axes represent the distances (1-corr) of edges on the networks of PET and MRI, respectively. Each dot is an edge and the color of edge is shown in (d). The abbreviations F, P, T, S, O and C represent  frontal, parietal, temporal, subcortical, occipital, and cerebellar regions.  R and L represent right and left hemispheres. The black line represents $y=x.$ The distance from dots to $y=x$ is estimated and its histogram is plotted above each DD plot.}
\label{fig:ddplot}
\end{figure}

The DD plots of ADHD, ASD and CON are shown in Fig. \ref{fig:ddplot} (a-c).
In each plot, the horizontal and vertical axes represent the edge weights of PET and MRI, respectively. Each dot represents an edge and different edges have different color as shown in (d). 
The cross-correlations between distances of PET and MRI in ADHD, ASD and CON were 0.354, 0.205 and 0.194, respectively.
They were not statistically significant when compared to the cross-correlation of random networks obtained by the permutation method. 
We estimated the distance from dots to the black line $y=x$ and plotted the histogram of distance above the DD plot.   
The percentage of edges that have larger distance in metabolic connections than in morphological connections was $50.4\%,$ $54.5 \%,$ and $68.0 \%$ for ADHD, ASD and CON, respectively. 
The edges of CON ($68.0 \% $) was significantly large with the level $.001$ as compared to $5000$ randomly permuted data.  


\subsection{$\beta_{0}-$plot between PET and MRI} 

\begin{figure}[thp]
\centering
\includegraphics[width=1\linewidth]{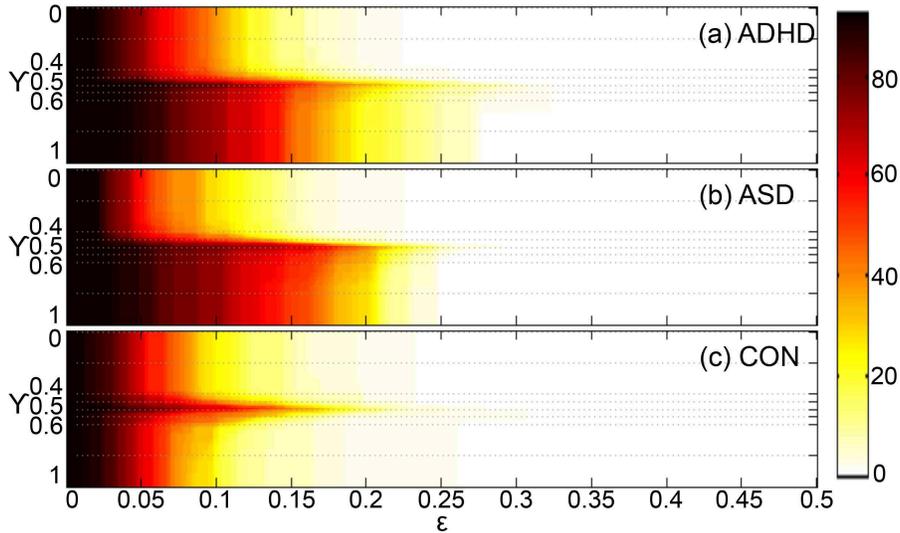}
\caption{$\beta_{0}$-plots between PET and MRI of (a) ADHD, (b) ASD, and (b) CON on the $\epsilon-\gamma$ plane. 
The vertical axis represents a mixing ratio $\gamma$ and the horizontal axis represents a new filtration value $\epsilon$ on the projection line with mixing ratio $\gamma.$
The color in $\beta_{0}$-plot in (a,b) is varied according to the zeroth Betti number $\beta_{0}$ between $1$ and the number of nodes $p=93.$
When $\beta_{0}$ becomes larger, the color changes from white to dark red as shown in the right colorbar.}
\label{fig:bettiplot}
\end{figure}

We applied the multifiltration method to two distance matrices of PET and MRI of each group.
The $\beta_{0}$-plots of ADHD, ASD and CON are illustrated in Figs. \ref{fig:bettiplot} (a), (b), and (c), respectively.
The horizontal and vertical axes represent the filtration value $\epsilon$ and the mixing ratio $\gamma$ between PET and MRI, respectively.
The color of $\beta_{0}$-plot is varied depending on the number of CCs $\beta_{0}$ at $(\epsilon,\gamma).$

We did KS-like test to show the difference of $\beta_{0}$-plot between ADHD and ASD, ADHD and CON, and ASD and CON using the permutation method.
The type-I errors for testing the difference between $\beta_{0}$-plots were $\alpha = 0.280$ (ADHD vs. ASD), $0.022$ (ADHD vs. CON), and $0.006$ (ASD vs. CON). 
Thus, in the view of multi-filtration, ADHD and CON, and ASD and CON were significantly different with the level $.05,$ but ADHD and ASD were not significantly different. 
The symmetry with respect to $\gamma=0.5$ on the $\epsilon-\gamma$ plane was $5.278,$ $6.393,$ and $0.500$ for ADHD, ASD and CON. 
When we tested the symmetry of ADHD, ASD and CON with $5000$ randomly permuted data, CON was statistically symmetric with the level $.005,$ but ADHD and ASD was not symmetric. 
Especially, ASD was more asymmetric than ADHD. 

\subsection{$\beta_{0}-$plot between groups} 

\begin{figure}[thp]
\centering
\includegraphics[width=1\linewidth]{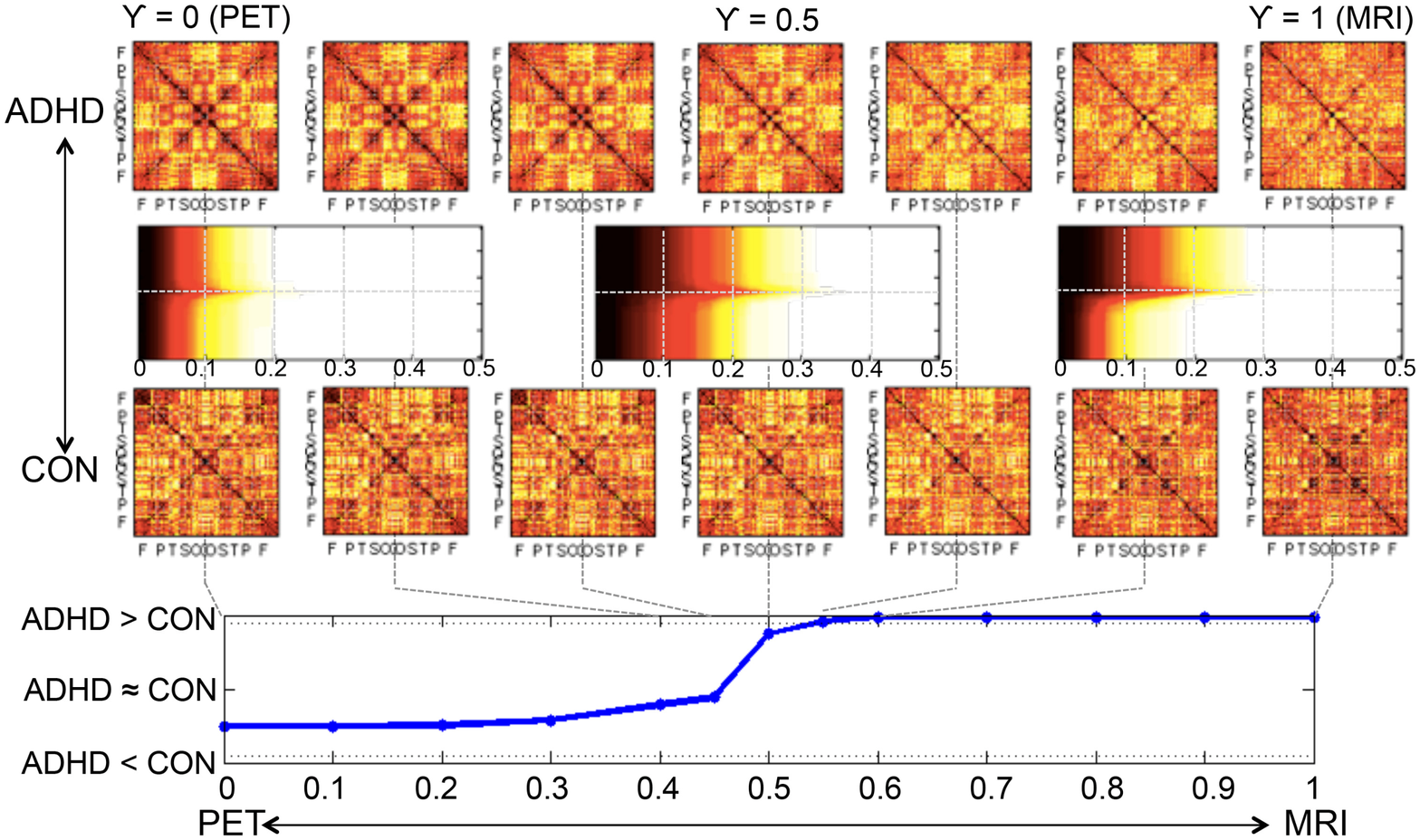}
\caption{\small{ADHD vs. CON. The first and third row show the sequence of integrated edge weight matrices of ADHD at $\gamma = 0,0.4,0.45,0.5,0.55,0.6,1.$ The second row represents the $\beta_{0}-$plot between ADHD and CON at $\gamma=0,0.5,1.$ The last row shows the p-value of the symmetry of $\beta_{0}-$plot in the second row. In the view of symmetry of $\beta_{0}-$plot, ADHD network shows slower connected pattern than CON network in the intervals of $\gamma=[0.55,1]$ with the level $.05.$}}
\label{fig:adhdvscon}
\centering
\includegraphics[width=1\linewidth]{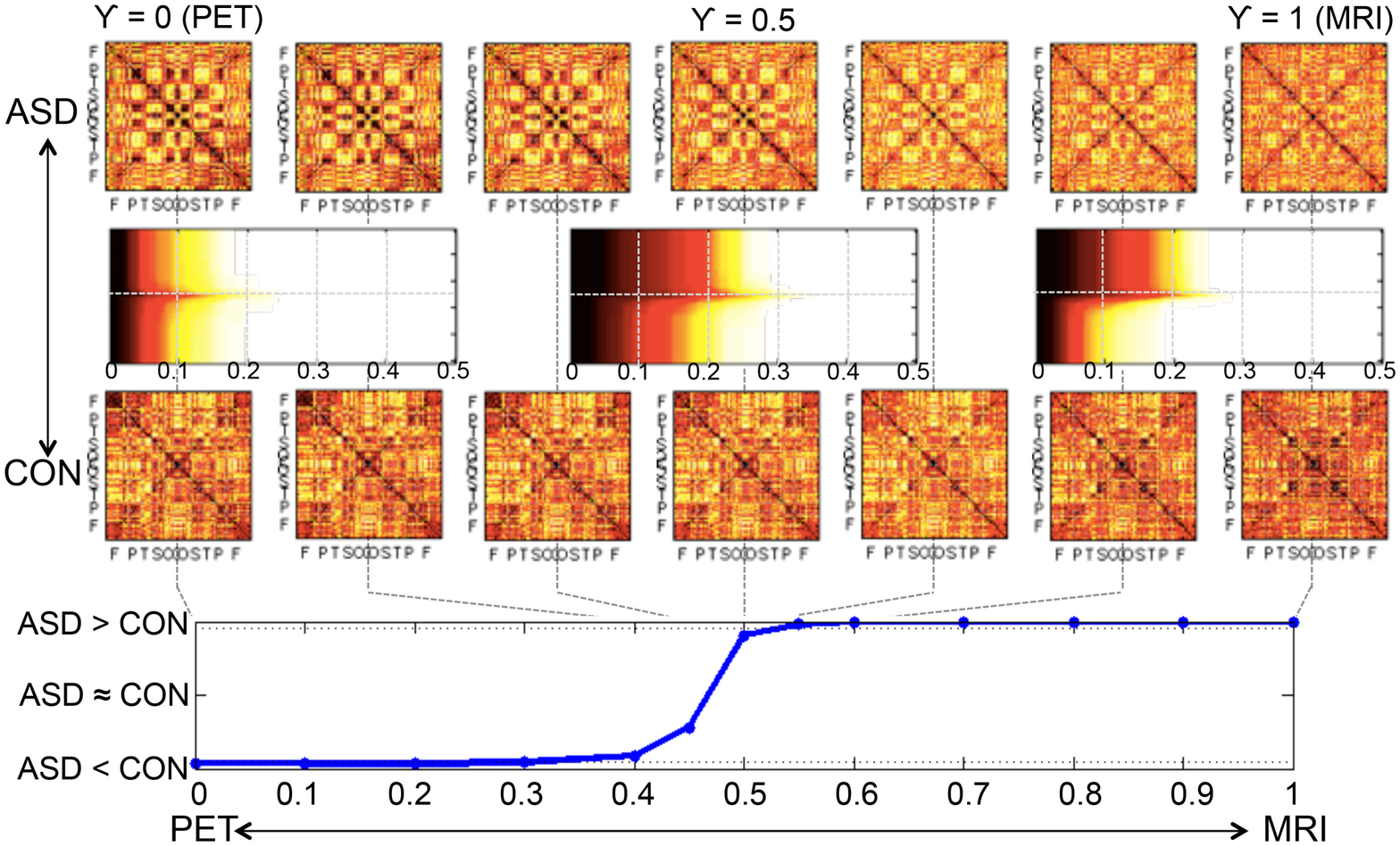}
\caption{\small{ASD vs. CON. In the view of symmetry of $\beta_{0}-$plot, ASD network shows faster connected pattern than CON network in the intervals of $\gamma=[0,0.3],$ but slower connected pattern in $\gamma=[0.55,1]$ with the level $.05.$}}
\label{fig:asdvscon}
\end{figure}

We estimated the sequence of integrated edge weight matrices of ADHD, ASD and CON at $\gamma = 0,$ $0.01,$ $0.02,$ \dots, $0.99,$ and $1.$ 
To test the difference between groups, we applied the $\beta_{0}-$plot method to the integrated edge weight matrices of pairs of groups at each $\gamma$ and estimated the symmetry of $\beta_{0}-$plot between ADHD and ASD, ADHD and CON, and ASD and CON at each $\gamma.$ 
The results are shown in Figs. \ref{fig:adhdvscon} and \ref{fig:asdvscon}.
In each figure, the first and third rows are the sequence of integrated edge weight matrices obtained by one-dimensional projection. 
Three $\beta_{0}-$plots between ADHD and CON or ASD and CON at $\gamma=0,0.5,$ and $1$ are shown in the second row.  
The last row shows the p-value of the symmetry of $\beta_{0}-$plots in the second row as compared to $5000$ randomly permuted data. 
The vertical and horizontal axes represent the p-value of the symmetry and the mixing ratio $\gamma$ of PET and MRI, respectively. 
There was no significant difference between ADHD and ASD in the view of global connected patterns. 
However, the difference of global connected pattern between ADHD and CON was found in the intervals of $\gamma=[0.55,1]$ where the information of MRI is mainly affected to the integrated edge weight matrix. 
It means that the brain regions of ADHD network were slowly merged compared to CON network. 
 The difference between ASD and CON was found in the intervals of $\gamma=[0,0.3]$ and $\gamma=[0.55,1].$ 
While the ASD network showed faster connected structure than CON network when the metabolic connectivity was mainly considered, it had slower connected structure than CON network when the morphological connectivity was mainly considered. 

\subsection{One-dimensional projection} 

\begin{figure}[thp]
\centering
\includegraphics[width=1\linewidth]{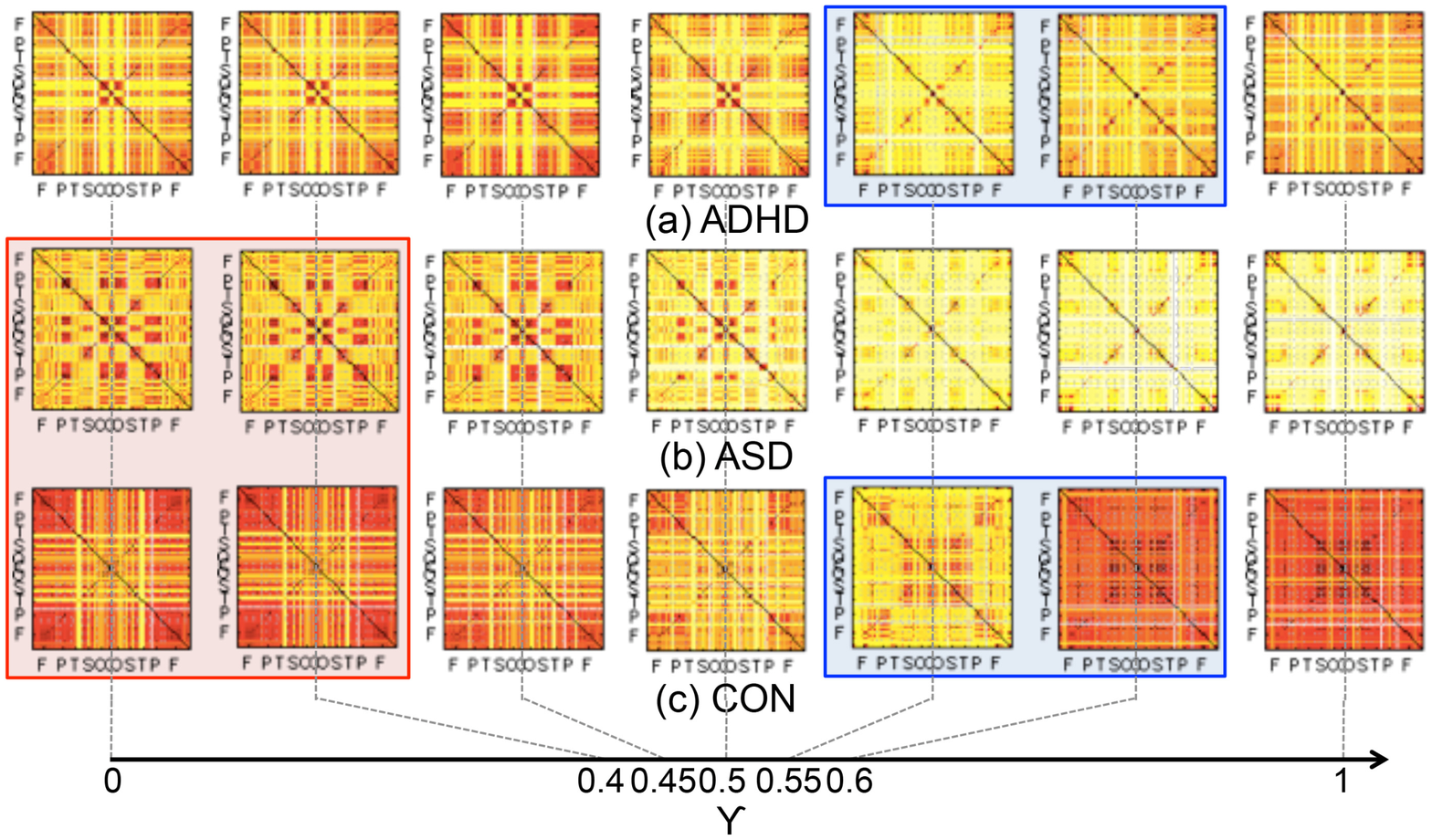}
\caption{\small{The sequence of integrated single linkage matrices of (a) ADHD, (b) ASD, and (c) CON. The mixing ratio $\gamma$ is $0$ (PET), $0.4, 0.45, 0.5, 0.55, 0.6,$ and $1$ (MRI). The SLMs of ADHD and CON are significantly different with the level $0.05$ in $\gamma = [0.54,0.68]$ in the blue box. The matrices of ASD and CON are significantly different in $\gamma=[0,0.40]$ in the red box.}}
\label{fig:d_petmri}
\end{figure}

We estimated the sequence of edge weight matrices and SLMs at $\gamma = 0,$ $0.01,$ $0.02,$ \dots, $0.99,$ and $1.$ 
The sequences of integrated edge weight matrices are plotted in the first and third rows of Figs.\ref{fig:adhdvscon} and \ref{fig:asdvscon}. The sequence of SLMs are are plotted in Fig. \ref{fig:d_petmri}.  
When we tested the difference of integrated edge weight matrices between ADHD and ASD, ADHD and CON, and ASD and CON based on GH distance and permutation method, ADHD and ASD were significantly different in the intervals of $\gamma = [0,0.46],$ and  ADHD and CON were significantly different in the intervals of $\gamma = [0.73,1],$ and ASD and CON were significantly different in the intervals of $\gamma = [0,0.19] $ and $[0.44,0.48]$ with the level $.05.$ 
When we tested the difference of SLMs between ADHD and ASD, ADHD and CON, and ASD and CON,  ADHD and CON were significantly different in $\gamma = [0.54,0.68]$ and ASD and CON were significantly different in $\gamma=[0,0.40]$ with the level $.05.$


\section{Discussion}

\subsection{Edge weights in respective metabolic and morphological connectivity networks}

Although the relationship 
between functional connectivity based on functional MRI and structural connectivity on DTI
has been reported in the literature \citep{bowman.2012.ni,greicius.2009.cc,skudlarski.2008.ni,vanheuvel.2009.hbm}, the relationship between metabolic and morphological connectivity based on FDG PET and T1 weighted MRI has
been rarely documented. 
Previous studies comparing functional and structural brain connectivity disclosed that the brain regions might be functionally connected through direct or indirect anatomical connections \citep{bowman.2012.ni}.
If metabolic connectivity is related closely to 
functional connectivity and morphological connectivity to 
structural connectivity, metabolic connections will tend to have larger distance than morphological connections. 
In experiments, the metabolic connectivity in CON showed larger edge weights than the morphological connectivity ($p=0.001$). 
However, such a result was not found in ADHD and ASD.

\subsection{Asymmetric change of ASD in the multidimensional $\beta_{0}$-plot}

In Fig. \ref{fig:bettiplot} (c), the changes of $\beta_{0}$-plot during filtration were almost symmetric between PET and MRI in control subjects.
However, the changes looked asymmetric in ASD between PET and MRI in Figs. \ref{fig:bettiplot} (b) as $\beta_{0}$ decreased quickly during filtration in metabolic connectivity of FDG PET but slowly in morphological connectivity of T1 MRI.
Neurodevelopmental disorders such as ASD involve 
abnormal functional and structural organizations rather than neuronal cell death or tissue loss 
as demonstrated by histo-pathological examinations in ASD
which showed that 
neuronal elimination decreased and myelination increased without neuronal population loss \citep{aylward.2002.neurology,courchesne.2001.neurology}.
This dysmaturation could have affected 
morphological connectivity, which we revealed in this investigation, and would have resulted in the slower merging of morphological connections between brain areas in Fig. \ref{fig:asdvscon}.
We interpreted this fact as an indication that the inter-regional connections were weaker.
In ASD, the metabolic connectivity showed the contrary finding of more rapid merging of metabolic connections between brain areas, which would represent
a compensatory effort of brain areas to overcome morphological loose connectivity.
In brief, the dysmaturation differentially affected the morphological association of gray matter in brain regions and the functional association of cerebral metabolic activity.

\subsection{Connectivities of difference between ADHD and controls in local connected components} 

When the global connected structures were compared by the symmetry of $\beta_{0}-$plot between groups in Fig. \ref{fig:adhdvscon}, it was shown that the information of MRI was more useful to discriminate ADHD and controls than one of PET.  
It seems to be related to the brain maturation delay in children with ADHD \citep{shaw.2007.pnas,vaidya.2012.ctbn}.
SLMs obtained from the integrated edge weight matrices between ADHD and controls were different in the interval of $\gamma=[0.54,0.68]$ with the level $.05.$ 
Using the integrated multimodal network at $\gamma=0.6,$ the different single linkage connections were found (a) between right supplementary motor area and other brain regions, (b) basal ganglia and paracentral lobule, temporal pole, amygdala, hippocampus, and occipital and cerebellar regions, and (c) right supramarginal gyrus in the parietal region and some frontal areas such as olfactory cortex, gyrus rectus, inferior frontal and superior frontal regions, (d) left Rolandic operculum in the frontal lobe and other brain regions, and (e) left Heschl gyrus in the temporal lobe and other brain regions 
($p < .05,$ uncorrected).  
The regions (a-c) in ADHD were connected at larger (single linkage) distances than in CON and regions (d,e) are opposite. 
The previous structural MRI studies showed that the children with ADHD have structural abnormalities in premotor cortex, basal ganglia structures, and cerebellar lobules \citep{shaw.2007.pnas,vaidya.2012.ctbn}. 
The resting-state functional connectivity studies have also shown that the frontal-striatal-cerebellar networks are weaker in children with ADHD  \citep{cao.2006.nr}.


\subsection{Connectivities of difference between ASD and controls in local connected components}

In the $\beta_{0}-$plot, the brain regions of ASD were rapidly merged in the metabolic connectivity, but slowly merged in the morphological connectivity, when we compared to controls.  
Our experiments had a tendency that the reduced local connections in ASD were mainly observed in the metabolic connectivity, but the increased local connections were in the metabolic connectivity. 
For example, the reduced connectivity in PET was found (a) between the occipital region and frontal and parietal regions, (b) in the visual cortex, (c) between left angular gyrus in the parietal regions and other brain regions, (d) between thalamus, amygdala, and hippocampus, (e) between thalamus and cerebellum, and (f) between left temporal pole and some frontal areas (gyrus rectus, middle frontal and superior frontal gyrus) and anterior cingulate cortex ($p < .05,$ uncorrected).  
On the contrary, the increased connectivity in MRI was mainly found (a) between some fronto-parietal regions, (b) basal ganglia and middle and inferior temporal gyrus, and (c) left inferior frontal gyrus and other brain regions ($p < .05,$ uncorrected). 
The abnormality of
the left perisylvian network for language could influence on language impairment and
be the cause of the socio-communication deficit of ASD \citep{defosse.2004.anneuro,just.2004.brain,knaus.2010.bl,baduell.2005.rn}.
The abnormal global processing along the dorsal visual pathway was reported in autism
as being related to `weak central coherence,' which complied with `reduced connectivity in
the visual cortex' \citep{pellicano.2005.np}.
The abnormalities of cerebello-thalamic circuitry and fronto-parietal connections were also reported in autism \citep{minshew.2010.con,takarae.2007.pr}.

\section{Conclusions}

By extending the previous filtration method to
a multimodal analysis, we
can observe the change of topological shape of multimodal brain connectivity by varying two different thresholds simultaneously.
We can take a quick look at the relationship between multiple imaging modalities using the multidimensional $\beta_{0}$-plot and find the group difference in it.
Whereas the controls showed relatively symmetric changes of connected structures between PET and MRI in the $\beta_{0}$-plot  during multidimensional filtration, ASD showed the more rapid changes of number of CCs at smaller threshold in metabolic connectivity, but the slower changes over a long range of the thresholds during filtration in morphological connectivity.
Although we used the number of connected networks as a topological invariant here, we suggest that the $\beta_{0}$-plot can be applied to the other graph theoretic measures such as small-worldness and betweenness centrality when 
the goal is to see their change when two different modalities are integrated.
By using a one-dimensional projection, we can integrate multimodal networks of PET and MRI at various mixing ratios and observe the change 
in topology of
the integrated network
in which 
PET and MRI information is mixed.
From the integrated network, we found that ADHD had increased morphological connectivity, and ASD also had increased morphological connectivity, but decreased metabolic connectivity. 
These results provide a multidimensional and multiscale homological understanding of disease-related 
metabolic and morphological networks. 

\section{Acknowledgments}
This research was supported by Basic Science Research Program through the National Research Foundation of Korea (NRF) funded by the Ministry of Education (0411-20150045), by the NRF grant funded by the Korea government (MEST) (No. 2011-003081), by the Original Technology Research Program for Brain Science through NRF funded by the Ministry of Education, Science and Technology (NRF-2015M3C7A1028926), and by NIH grant UL1TR000427 and Vilas Associate Award from Univ. of Wisconsin.

\bibliographystyle{abbrvnat}
\bibliography{leehk}

\end{document}